\documentclass{phbauth}
\usepackage{graphicx}

\begin{document}
\begin{frontmatter}

\title{Onset of superconductivity in 2D granular films composed of Bi clusters}

\author{N. Chandrasekhar\thanksref{thank1}},

\address{Department of Physics, Indian Institute of Science,
Bangalore 560 012, India}

\thanks[thank1]{E-mail: chandra@physics.iisc.ernet.in}
\begin{abstract}
Two dimensional films comprising of Bi clusters ranging in size from 25 to 100 $\mathrm{\AA}$ show an influence of the underlying matrix on normal state resistivity and the superconducting transition temperature $T_\mathrm{c}$.  This is in contrast to crystalline Bi.  Quantum size effects are observed by changing the deposition temperature,
which determines cluster size.  These observations can be 
attributed to superconductivity at the surface.
Implications for the requirement of a critical number of
superconducting clusters before the establishment of global phase coherence are briefly discussed.
\end{abstract}

\end{frontmatter}

\section{Introduction}

Insulator to metal/superconductor
transitions, onset of conductivity and nature of transport
have been explored using quench condensed thin films.  Microstructure and
morphology govern these phenomena, which depend on how the films are grown.
\cite{buck,abel,stro,havi}  The processes controlling the onset of
conduction in such films remain controversial.  The prime cause for this
is the lack of direct structural information.  The conventional wisdom is that at the onset of conductivity, film is
composed of islands.  The opposing school of thought questions this
since thermally activated mechanisms for island growth freeze out
at cryogenic temperatures, with adatoms sticking wherever they land.\cite{dani}   

Of two recent experiments to probe the onset of conductivity,
\cite{ekinc,henn} one study probes the IR 
conductivity, where $\sigma$ is found to increase with $\omega$ but in 
disagreement with predictions of weak localization.  
The data are consistent with a percolation type model.
In contrast the other experiment probes the structure with an STM
and finds that even at 75\% coverage, the films are not conducting.
This latter experiment is inconsistent with a percolation model.

Several materials (eg. Bi, Al) exhibit an increase in the transition temperature when
quench condensed as disordered thin films.  Two different mechanisms have
been proposed for this, with experiments yielding
ambiguous results.  One mechanism attributes the
enhancement to softening of the phonon modes, caused
by the increased surface to volume ratio for nanoparticles.\cite{dick} 
The opposing interpretation is an enhancement
of the $T_\mathrm{c}$ by a screening mechanism which results from proximity
to a dielectric.\cite{hura}  Our experiments indicate that such an effect
plays an important role, as shown in Fig. 1.\cite{sam}
Such results have been reported earlier.\cite{mood,cjad}

Experimental evidence from
both real space STM and reciprocal space RHEED studies indicates the presence of
clusters.\cite{ekinc,sam}  Neither technique
has been fully exploited to resolve the controversy over the 
crystalline/amorphous nature of these.  Small size implies discreteness of the
energy levels, whose spacing may be large compared to thermal energy kT,
making the assumption of metallic behavior incorrect.\cite{giav}

The nature of disorder would be different in films with Ge underlayers,
since Ge offers dangling bonds.  
Since the resistances of films with underlayers
are higher, we believe that this is evidence of an interplay between
screening and localization.  Such an effect is not necessarily small,
being of the order of typical logarithmic correction,
$\mathrm{e^2/2\pi^2\hbar}$.
Thus the underlayers play a twofold role.  Observations of a metallic monolayer
increasing the transition temperature are relevant here.\cite{shapo}
Further, experimental observations
that slight thickness variations change a film from an insulator to a 
superconductor, indicate a role for percolation.  The two contradictory
experiments discussed above may then be reconciled as under.\cite{ekinc,henn}

Onset of superconductivity may then be
regarded as a consequence of varying the percentage coverage of Josephson
like bonds.  Percolation threshold corresponds to the onset of macroscopic
phase coherence with 50\% of the junctions among superconducting clusters being phase coherent.  Due to quantum size effects not all clusters will be superconducting.  Hence, although 75\% of the substrate may be
covered by material as evidenced by STM observations, there may be no
conductivity across the film, implying that a critical
number of superconducting clusters (that yield 50\% phase coherent junctions) are a must for establishment of global
phase coherence.  A more detailed analysis of this will be published elsewhere.

\begin{figure}[btp]
\begin{center}\leavevmode
\includegraphics[width=0.9\linewidth, height=0.9\linewidth]{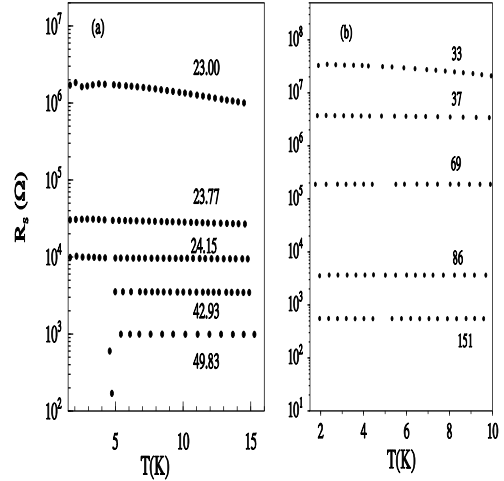}
\caption{ 
Evolution of resistivity vs temperature for Bi films on (a) 10$\mathrm{\AA}$ Ge underlayer (b) solid xenon.  The Ge was deposited on amorphous quartz at room temperature
and the Bi deposited on both at 1.5 K. The numbers on the right hand side indicates nominal thickness in $\rm \AA$.
}\label{figurename}\end{center}\end{figure}

\begin{ack}
This work was supported by DST, Government of India.
\end{ack}

\end{document}